\documentclass[fleqn,preprintnumbers,amsmath,amssymb,prb]{revtex4}
\usepackage{graphicx}
\usepackage{dcolumn}
\usepackage{bm}
\usepackage{epsf}
\def\beq{\begin{equation}}
\def\eeq{\end{equation}}
\def\beqar{\begin{eqnarray}}
\def\bdm{\begin{displaymath}}
\def\beqas{\begin{eqnarray*}}
\def\edm{\end{displaymath}}
\def\eeqar{\end{eqnarray}}
\def\eeqas{\end{eqnarray*}}

\def\lbr{\langle}
\def\rbr{\rangle}
\def\eps{\epsilon}
\def\vareps{\varepsilon}
\def\tm{\times}

\def\bn{\bar{n}}

\def\hb{\hbar}
\def\btu{\bigtriangleup}
\def\lam{\lambda}
\def\om{\omega}
\def\Del{\Delta}
\def\del{\delta}

\def\Gam{\Gamma}

\def\kap{\kappa}
\def\alp{\alpha}

\def\kp{k^{\prime}}
\def\kdp{k^{\prime \prime}}
\def\lp{l^{\prime}}
\def\ldp{l^{\prime \prime}}

\def\non{\nonumber}

\begin{document}
\title{Equation of state for asymmetric nuclear matter \\
with infinite-order summation of ring diagrams}
\author{J. Shamanna}
\email{jaya@vbphysics.net.in}
\affiliation{Physics Department, Visva Bharati University, 
Santiniketan 731235, India}
\author{T.T.S. Kuo}
\affiliation{Physics Department, State University of New York at Stony Brook
Stony Brook, NY 11794-3800, USA}
\author{I. Bombaci}
\affiliation{Dipartimento di Fisica, Universita di Pisa, Italy}
\author{Subhankar Ray}
\affiliation{Dept. of Physics, Jadavpur University, Calcutta 700032, India}
\date{18 August 2005}
\begin{abstract}
The particle-particle hole-hole ring-diagram summation method 
is employed to obtain the equation of state
of asymmetric nuclear matter over a wide range of asymmetry fraction.
Compared with Brueckner Hartree-Fock and
model-space Brueckner Hartree-Fock calculations, this approach gives a 
softer equation of state, increased symmetry energy and a lower value for the 
incompressibility modulus which agrees quite well with the values
used in the hydrodynamical model for the supernovae explosion.
\end{abstract}
\maketitle
\setcounter{equation}{0}
\setcounter{page}{1}

\section{Introduction}
A primary aim of microscopic nuclear theories is to derive
the various nuclear properties such as the binding energy per nucleon
$(BE/A)$ and saturation density $(\rho_0)$ of nuclear matter, starting
from fundamental nucleon-nucleon (NN) interactions. The well-known BHF
approach is one such standard nuclear matter theory.
In terms of G-matrix diagrams, the BHF theory is, however, only
a lowest-order approximation; the ground-state energy shift
$\Delta E_0$ for nuclear matter, due to the NN interaction,
 is  given merely by the
first-order G-matrix diagram, fig. 1(a), namely
\beq\label{ebhf}
\Del E_0^{BHF}=\sum _{ab} n_a n_b\lbr ab\vert G^{BHF}(\om =
             \eps_a + \eps_b )\vert ab \rbr
\eeq
In the above the $n$'s are the unperturbed Fermi-Dirac
distribution functions,
$n_k$=1 if $k\leq k_F$ and =0 if $k>k_F$. $k_F$ is the Fermi momentum.

\begin{figure}[h]
\resizebox{!}{2in}
{\includegraphics{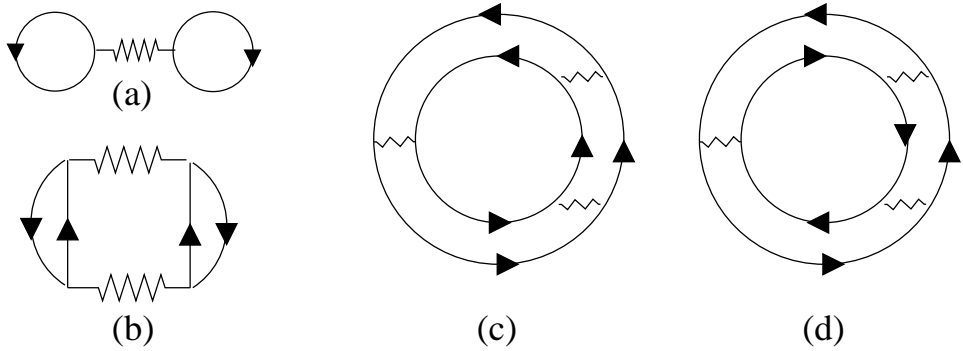}}
\caption{Some lower order ring diagrams}
\label{lowring}
\end{figure}

The  single-particle (s.p.) energies denoted
by $\eps$, are determined self-consistently using the BHF theory.
As is well known, this approach has not been very successful; it has, in
general, not been able to simultaneously reproduce the binding energy 
per nucleon ($BE/A=-16\pm 1$ MeV) and the saturation  Fermi momentum
($k_F^{sat}=1.35\pm 0.05$ fm$^{-1}$).
When plotted on a energy-density plane, results of various BHF
calculations
for nuclear matter invariably lie, more or less, on a band, the Coester band
\cite{ccdv70}, which significantly misses the "experimental (empirical) box"
for $BE/A$ and $k_F^{sat}$.

The ground state of nuclear matter in BHF theories is
treated as merely a Fermi sea.
Particle-hole fluctuations near the Fermi sea which represent the
long-range correlations are not considered. It should be
of interest to allow for such Fermi-sea fluctuations, and they may be important
in determining the stiffness of the nuclear-matter equation of state (EOS).
Yang, Heyer and Kuo\cite{yhk86} proposed an elegant and rigorous 
method for summing
up particle-particle hole-hole (pphh) and particel-hole (ph) ring 
diagrams to all orders
for the calculation of ground state energies of general many-body
system. Inclusion of these classes of diagrams to all orders takes into
account the Fermi sea fluctuations and long-range nuclear correlations.

With this motivation, several calculations for symmetric
nuclear matter have been carried out with the inclusion of certain
class of ring diagrams to all orders\cite{syk87,jkm88,jmk89}.

In comparison with conventional BHF \cite{b71,bbn85} calculations of
nuclear matter, the inclusion of the particle-particle hole-hole (pphh)
ring diagrams to all orders has the desirable effect of both increasing
the nuclear matter binding energy
and lowering its saturation density.
The final expression for the ground-state energy shift
in the pphh ring diagram summation with a model-space
reaction G-matrix ($G^M$) is given as
\beq\label{ering}
\Del E^{ring}_0= \int^1_0 \frac{d\lam}{\lam}
{\rm tr}\{Y_m(\lam) Y^{*}_m(\lam)G^M(\om=\Del^{A-2}_m(\lam))
\lam \}
\eeq
Comparing with the corresponding BHF result of eq. (\ref{ebhf}), the main
difference between the two methods is the replacement of the unperturbed
occupation factors $n$, BHF G-matrix $G^{BHF}$ and  starting energy
($\epsilon_a+ \epsilon_b$) in the BHF expression
by the RPA amplitudes $Y$, model-space G-matrix $G^M$ and RPA energies
$\Delta$, respectively. In the above $\lambda$ is a strength parameter
introduced to facilitate the calculation, as will be discussed later.

In the present paper, we would like to extend the above ring-diagram
scheme to asymmetric nuclear matter, which is in several ways of more physical
importance than symmetric (N=Z) nuclear matter.
The study of the EOS of asymmetric nuclear
matter has become, in the last few years, a subject of renewed interest
particularly in connection with astrophysical problems\cite{bck85,bgh85},
such as supernovae explosions and the evolution of
neutron stars. For these physical processes, the nuclear matter involved
is predominantly not symmetric, and it is the EOS of asymmetric
nuclear matter (with a large neutron excess) which plays an important role
for them. Furthermore, the nuclear matter probed by heavy-ion experiments
is also generally asymmetric. 
Here the proton-neutron
ratio is about 2/3 for both the target and the projectile, and thus 
the resulting nuclear matter is likely to be asymmetric with the same 
proton-neutron ratio.
The range of densities sampled by astrophysical systems such as
the supernovae and the neutron star vary over an appreciably wide range 
as do their isospin asymmetries. 

According to the model of prompt explosion\cite{k89}, which has been
widely employed to explain the explosion mechanism of a supernova, the
nuclear-matter EOS needs
to be relatively soft\cite{bck85} for the explosion to take place.
It should be of interest to investigate the effect of ring diagrams on the
stiffness of asymmetric nuclear matter, as we shall carry out later.
To our knowledge, such investigations have not yet been performed.

Asymmetric nuclear matter calculations have been done using
the Skyrme interactions\cite{jmz83,sylk86}, the Gogny
interaction\cite{szs90} and using the Brueckner-Bethe-Goldstone
(BBG) approach\cite{bl91,bkl94}.
Our present calculation is a continuation of the model-space BHF (MBHF)
calculations for asymmetric nuclear matter carried out by
 Song, Wang and Kuo\cite{swk92}.
In the past, the asymmetric nuclear matter properties
were often extracted by interpolating the two extreme cases of
symmetric and pure
neutron matter, with an empirical parabolic approximation\cite{cdl87,wff88}.
The validity of this empirical practice seems to have not been investigated,
and we would like to carry out such an investigation
 by carrying out a sequence of
ring-diagram nuclear matter calculations, covering a wide range 
of proton-neutron ratios and  baryon densities.

\section{Formalism}
Asymmetric nuclear matter is a system consisting of
$N$ neutrons an $Z$ protons with $N \neq Z$.
For symmetric nuclear matter $N$ and $Z$ are identical with the same 
Fermi momenta. For asymmetric matter however, the neutrons and protons 
are treated as non-identical particles with different Fermi mommenta.

We introduce a parameter $\alp$
 as a measure of the asymmetry in nuclear matter, namely
\beq\label{alp}
\alp = \frac{(\rho_n -\rho_p)}{\rho}=\frac{(N-Z)}{A}
\eeq
\noindent where $\rho$, $\rho_n$ and $\rho_p$ are respectively
the nuclear, the neutron and the proton densities.
The neutron and the proton Fermi momentum are
\beq\label{mom}
k_F^n={(3 \pi^2 \rho_n)}^{1/3}~~~~~~~~~~
k_F^p={(3 \pi^2 \rho_p)}^{1/3}
\eeq
 An average Fermi momentum is defined as
\beq\label{avmom}
k_F={(\frac{3}{2} \pi^2 \rho)}^{1/3}
\eeq

\section{Model-space G-matrix and single-particle spectrum}
As with the symmetric case we start with a Hamiltonian $H=T+V$.
Introducing a single particle (s. p.)
potential $U$ we rewrite it as $H=(T+U)+(V-U)$. $V$ is the NN potential
such as the Paris\cite{parispot} or the Bonn\cite{bonnpot} potential.
A model space P is defined as a configuration space where all
nucleons are restricted to have the momentum $k \leq k_M$, $k_M$
being the momentum space boundary of P.
A typical value for $k_M$ is $2k_F$ where $k_F$ is the Fermi momentum.
As we shall discuss later, most of the calculations in the present work
are performed using $k_M$=3.0 fm$^{-1}$.

Similar to the case of symmetric nuclear matter we use a model-space
Hartree-Fock method \cite{mk83,kmw84,kmm86}
to determine $U$. This leads to the following self-consistent equations
for the s.p. spectrum $\eps_i ^M$, namely
\beqar\label{spec}
\eps^{M}_{i}&=&t_{k_{i}}+ \Gam_{k_i}(k_{i},\tau_{i}) \\
\Gam_{k_{i}} (\om, \tau_{i}) &=& 2 \sum_{ \stackrel{\tau_j, s_i, s_j} 
{k_j \leq k_F^{\tau_j}}} \lbr k_{i} k_j |G^{M}(\om+\eps_j)| k_{i} k_j
\rbr    \;\;\;\;\;\;\;\;\; k_{i} \leq k_M      \\
\Gam_{k_{i}} (\om,\tau_i) & = & 0
\;\;\;\;\;\;\;\;\;\;\;\;\;\;\;\;\;\; k_{i} > k_M
\eeqar
Note that the subscript $i$ represents both momentum and isospin, namely
$i \equiv ({\bf k}_i, \tau_i)$ with
{\bf k}$_i$ denoting the momentum and $\tau_i$ the $z$
component of the isospin of the $i$th nucleon.
The single particle kinetic energy is $t_{k_{i}}=\hbar ^{2} k_{i}^{2}/2 m$,
$G^{M}$ is the reaction matrix to be specified later. The Fermi momentum
is represented by $k_F ^{\tau_j}$ with neutron Fermi momentum $k_F ^n$ 
for $\tau_j=-\frac{1}{2}$, and proton
Fermi momentum $k_F ^p$ for $\tau_j=\frac{1}{2}$. The model-space momentum 
boundary is $k_M$ and is taken to be the same for neutrons and protons.
The s.p. potential is the one-body vertex
function $\Gam $ evaluated at the self-consistent energy $\om=\eps_{k_{i}}$
\beq\label{sppot}
U(k_{i},\tau_i)= \Gam_{k_{i}}(\eps_{k_{i}},\tau_i)~, ~~~~~~~~~~i\equiv n,p
\eeq
$U(k_{i},\tau_i)$ is determined self-consistently for $k_{i} \leq k_M$, and for
$k_{i} > k_M $ we set $U=0$. We also use an effective mass description
for the single particle spectrum as
\beq\label{effm}
\eps^M_{k_i} = \left\{ \begin{array}{ll}
  (\hbar^2/2m_q^*) k_i^2+\btu_q, & \;\;\;\;\; k_i \leq k_M  \\
  (\hbar^2/2m) k_i^2, & \;\;\;\;\;  k >k_M 
  \end{array} \right. 
\eeq
with four parameters $m^{*}_q, \btu_q $, $(q=n,p)$.
The effective mass $m^{*}$ and the zero point energy $\btu$ are
determined self-consistently.

The model-space reaction matrix $G^{M}$
satisfies the Bethe-Goldstone equation
\beq\label{bg}
\lbr i j|G^{M}(\om)|m n \rbr = \lbr ij|V|mn \rbr +\sum_{k,l}
\frac{ \lbr ij|V|kl \rbr Q^{M}(k,l) \lbr kl|G^{M}(\om)|mn \rbr }
{\om-\eps_k -\eps_l} 
\eeq
In the above equation $i,j,m,n,k$ and $l$ are single particle states, each with
momentum {\bf k}, isospin $\tau$.
$V$ is the nucleon-nucleon interaction. The energy variable $\om$ in 
the denominator of eq. (11) is given by
\beq\label{om}
\om = \eps_k ^{M}+\eps_l ^{M}
\eeq
The two particle correlations considered are those where at least one
particle is out of the model space. Hence the Pauli operator $Q^M$
is defined as
\beq\label{pauli}
Q^{M}(k,l) = \left\{ \begin{array}{ll}
 1, & \mbox{if max}(k_k,k_l) > k_M \;\; \mbox{and} \;\; k_k > 
k_F^{\tau_k}, k_l > k_F^{\tau_l} \\
 0, & \mbox{otherwise}
 \end{array} \right.
\eeq
Note that $Q^M$ is different for each of the following three cases 
of the intermediate two nucleon state.

\begin{figure}[h]
\resizebox{!}{2.0in}
{\includegraphics{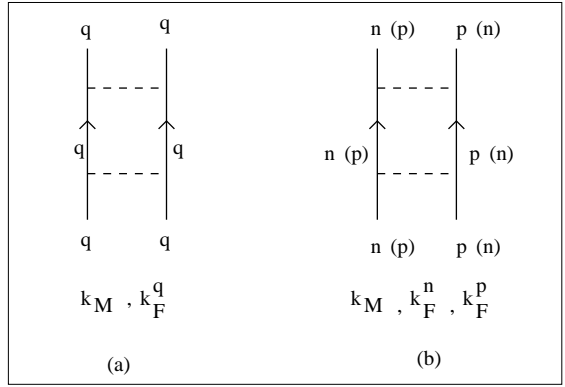}}
\caption{Intermediate two nucleon states in the asymmetric case \\ 
(a) $nn$ or $pp$ (b) $np$ or $pn$.}
\label{qnp}
\end{figure}

For the $nn$ ($pp$) case (fig.\ref{qnp}(a)), the intermediate nucleons
are identical, and only
$k_M$ and $k_F^q$ , $q=n~or~p$, enter the calculation. For
the $np$ or $pn$ case however, $k_M$, $k_F^n$ and $k_F^p$ all
play a role in determining $Q^M$ (fig.\ref{qnp}(b)).
It is convenient to carry out the above calculation in the relative
and center of mass (RCM) frame.
We choose our relative momentum {\bf k}
and center of mass momentum {\bf K} as
\beq\label{kK}
{\bf k}=\frac{1}{2}{\bf (k_{k}-k_{l})}\;,\;\;\;\;\;\;\; 
{\bf K}={\bf (k_{k}+k_{l})}
\eeq

\begin{figure}[h]
\resizebox{!}{2.0in}
{\includegraphics{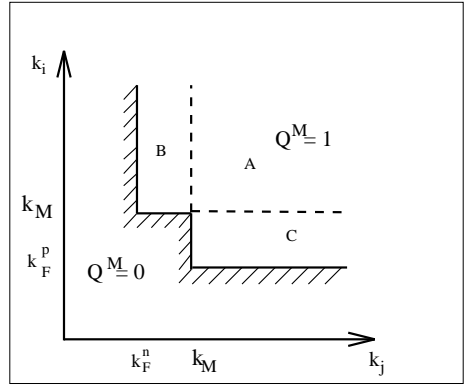}}
\caption{Pauli operator $Q^{M}(k_{k},k_{l})$}
\label{paul}
\end{figure}

\begin{figure}[h]
\resizebox{4.0in}{2.0in}
{\includegraphics{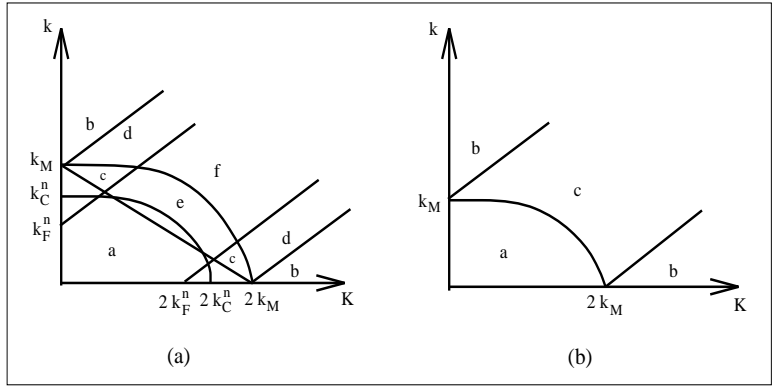}}
\caption{Angle averaged Pauli operator $\bar{Q}^{M}_B$ and $\bar{Q}^{M}_A$.}
\label{paul2}
\end{figure}

First we replace the Pauli exclusion operator $Q^{M}$, which is a
function of the laboratory momenta by its angle average approximation
$\bar{Q}^{M}$.
We divide the plane of the two laboratory momenta into three regions
A, B and C as shown in fig.\ref{paul}. The values of $Q^M$ in the 
three regions are
shown in the figure. Each of the regions is  transformed into
the RCM frame.
Then the angle-averaged value of $Q^M$ is
\beq\label{qbar}
\bar{Q}^{M}=\bar{Q}^{M}_{A}+\bar{Q}^{M}_{B}+\bar{Q}^{M}_{C}
\eeq
where, for region B, we have
\beq\label{qbarb}
\bar{Q}^{M}_B(k,K) = \left\{ \begin{array}{ll}
  0 & \mbox{regions} \;\; a,b   \\
  ((k+\frac{1}{4}K)^2-k_M ^2)/2kK & \mbox{region} \;\; c \\
  (k_M ^2-(k-\frac{1}{2}K)^2)/2kK & \mbox{region} \;\; d \\
  (2k^2+\frac{1}{2}K^2-k_M ^2-(k_F^n)^2)/2kK & \mbox{region} \;\; e  \\
  (k_M^2-(k_F^n)^2)/2kK & \mbox{region} \;\; f
  \end{array} \right.
\eeq
\noindent The subdomains a to e are shown in 
fig.\ref{paul2} and
$2(k_C^n)^2=[(k_F^n)^2+k_M^2]$.
Angle average approximations are standard (and indispensible) in
treating Pauli exclusion operators in nuclera matter calculations
and are generally considered to be accurate\cite{bethe71,sp72}. 
This technique has 
the advantage of making the model-space reaction matrix diagonal in the
RCM vriables.Recent studies of an exact pauli operator calculation have been 
reported\cite{smc99}. It is not very clear whether such  calculations
show an appreciable difference in the binding energy per nucleon
and the saturation density from previous studies.
Angle averages for region C may be obtained from the above 
by substituting $k_F ^p $ for $k_F ^n$.
The calculation above was an illustration of the case where the
intermediate state contains one neutron and one proton. The other
two possibilities may be readily obtained from above by suitable substitution
of the relevant Fermi momentum.

A similar RCM mapping of region A in fig.\ref{paul}
gives
\beq\label{qbara}
\bar{Q}^{M}_A(k,K) = \left\{ \begin{array}{ll}
 0 & \mbox{region} \;\; a   \\
 1 & \mbox{region} \;\; b   \\
 (k^2+\frac{1}{4}K^2-k_M^2)/kK & \mbox{region} \;\; c
  \end{array} \right.
\eeq
For symmetric nuclear matter $k_F^n=k_F^p$ so that
$\bar{Q}^M_B=\bar{Q}^M_C$ and the angle averaged $Q^M$
is the same as that given in eq. (\ref{qbarb})

Using the angle averaged Pauli operator the model space reaction matrix can
be decomposed into separate partial wave channels as
\beq\label{gmpw}
\lbr kl|G^{M}(\om,K w)|\kp \lp \rbr = \lbr kl|V|\kp \lp \rbr
+\frac{2}{\pi} \int_{0} ^{\infty} d \kdp~ {\kdp}^{2} 
\sum_{\ldp} \frac{ \lbr kl|V|\kdp \ldp \rbr \bar{Q}^{M}(\kdp,K) 
\lbr \kdp \ldp|G^{M}(\om,K w)|\kp \lp \rbr }{\om-H_0(\kdp,K)} 
\eeq
where $w$ denotes the partial wave quantum numbers $(l \lp SJT)$, and $K$
is the center of mass momentum. The $K$ and $(SJT)$ quantum numbers
associated with the bra and ket vectors have been suppressed for
simplicity. For example, $\lbr kl |$ should in fact be $ \lbr klSJT,K|$.
Our convention for plane waves is
\beq\label{plw}
\lbr {\bf r}|klSJ \rbr =j_l(kr) {\cal Y}_{lSJ}({\bf \hat{r}})
\eeq
where $j_l(kr)$ is the spherical Bessel function, and $\cal Y$ is the
vector spherical harmonics corresponding to {\bf l=S+J}. The angle averaged
reaction matrix $G^{M}(\om,K w)$ is diagonal in $K$ and $w$. This
is a consequence of using angle averages for the projection operator $Q^{M}$
and the energy denominator.

From eq. (\ref{om}) the energy variable in the laboratory frame is 
$\eps_i+\eps_h$.
Using RCM variables, the energy denominator $\om$ for the neutron spectrum 
is given by
\beq\label{nspec}
\om =\frac{\hb ^2}{m_p^*}k^2+\frac{\hb ^2}{4m_p^*}K^2+\btu_n 
+\btu_p+ \left[\frac{\hb ^2}{2m_n^*}-\frac{\hb ^2}{2m_p^*}
\right](k^n)^2
\eeq
Similar expression is obtained for the proton spectrum by replacing the
subscripts and superscript $n$ by $p$.
The other term in the denominator $H_0=\eps_m+\eps_n$ is the unperturbed energy
of the intermediate states and is also
angle dependent. The momentum variables $k_m$ and $k_n$ corresponding
to $k$ and $K$ may be in either of the three regions A, B or C.
In region A, both intermediate particles have momentum larger than
$k_M$. Therefore we have
\beq\label{ha}
H^{\rm A}_0(k,K)=\frac{\hb^2}{m}k^2+\frac{\hb^2}{4m}K^2
\eeq
In region B, we have a proton with momentum larger than $k_M$
and a neutron with momentum between $k_F^n$ and $k_M$.
 We use an angle average approximation for $k_m^2$
i.e., $\lbr k_m ^2 \rbr =\frac{1}{2}[(k_F ^n)^2+k_M ^2]$ and obtain
\beq\label{hb}
H^{\rm B}_0(k,K) = \frac{\hb^2}{m}k^2+\frac{\hb^2}{4m}K^2 + \btu_n  
+\frac{\hb^2}{4m}[\frac{m}{m^* _n} -1][(k_F ^n)^2+k_M ^2]
\eeq
The spectrum of $H_0$ for region C is the same as that of region B
with the subscripts and superscript $n$ changed to $p$.

\noindent Finally the single particle potential in the (RCM) frame for
$k_1 \leq k_F^{\tau_q}$ with $(q \equiv n,p)$ is given as
\beqar\label{rcmpot1}
U(k_{1},\tau_{q})&=& \frac{16}{\pi}\sum_{lSJ \tau_j}(2J+1) \int_{0}^{k_{-}}
dk~ k^2 G^M _{lSJ \tau_q \tau_j}(k,\bar{K}_1)  \non \\
& & +\frac{2}{\pi k_1} \sum_{lSJ \tau_j}(2J+1) \int_{k_{-}} ^{k_+}
dk~ k [(k_F ^{\tau_j})^2-k_1 ^2+4k(k_1-k)]
G^M_{lSJ \tau_q \tau_j}(k,\bar{K}_2) \non \\
\eeqar
And for $k_1 > k_F^{\tau_q}$ but less than $k_M$ we have
\beq\label{rcmpot2}
U(k_{1},\tau_{q})=\frac{2}{\pi k_1} \sum_{lSJ \tau_j}(2J+1) \int_{k_{-}} ^{k_+}
dk~ k [(k_F ^{\tau_j})^2-k_1 ^2+4k(k_1-k)]G^M_{lSJ \tau_q \tau_j}(k,\bar{K}_2)
\eeq
where
\beqar\label{kminus}
k_{-}&=&\frac{1}{2}|k_F ^{\tau_j}-k_1|  \non  \\
k_+  &=&\frac{1}{2}(k_F ^{\tau_j}+k_1)   \non  \\
\bar{K^2 _1} &=& 4(k_1 ^2+k^2)     \non  \\
\bar{K^2_2} &=& 4(k_1 ^2+k^2)-(2k+k_1-k_F ^{\tau_j})(2k+k_1+k_F ^{\tau_j})
\eeqar
and the partial wave reaction matrix elements are given by
\beq\label{pwrm}
G^M_{lSJ \tau_i \tau_j}(k,K)= \lbr kl \tau_i \tau_j|
G^M(\om,KlSJ)|kl \tau_i \tau_j \rbr
\eeq
Eq. (\ref{gmpw}) is in isospin representation with well defined 
total isospin $T$
of the two nucleon state. The reaction matrix elements in the neutron-proton
representation i.e, eq. (\ref{pwrm}), are related to those in the isospin
representation by Clebsch-Gordon coefficients; which explains the factor
of $2$ in the expression for the s.p. potential.

The s.p. potentials $U(k_1,\tau_n)$, $U(k_1,\tau_p)$ and the
spectra $\eps_{n}$, $\eps_{p}$ are calculated self-consistently as
described previously.

\noindent A main purpose here is to convert the strong $V$ interaction to a
well-behaved G-matrix interaction. In so doing, one must make sure that
there is no double counting. Thus a double-partitioned approach is adopted,
 treating the nucleon-nucleon correlations with
high momentum (i.e. those with $Q^M=1$)  within the $G^M$ matrix,
while taking care of  the low-momentum correlations by the
pphh ring diagrams.
We would like to express the energy shift in terms of the
model space G-matrix\cite{syk87}.

Proceeding as in Ref. (3) we define a model-space G-matrix
and formulate an expression for the energy shift in terms of
the transition amplitudes as
\beq\label{eshift}
\Del E^{pp}_0= \int^1_0 \frac{d\lam}{\lam} \sum_{\stackrel{m}{(A-2)}}
\sum_{\stackrel{i>j,k>l}{\in P}}Y_m(ij,\lam)
Y_m^{*}(kl,\lam)G^M_{klij}([\om=\Del^{A-2}_m(\om)])\lam
\eeq
\noindent with
\beqar\label{dely}
\sum_{e>f} \{ \del_{ij,ef} (\vareps_i+\vareps_j) +
(1-n_i-n_j) \lambda L_{ijef}(\om)\}Y_m(ef,\lam) &=&
\Del^{A-2}_m(\om)Y_m(ij,\lam)~ \non  \\
\om &=&\Del^{A-2}_m(\om)
\eeqar
For the neutron-neutron ($nn$) or the proton-proton ($pp$) cases,
the inequality signs for the summation indices $i,~j,~k,~l$ restrict the
momentum $k_i \geq k_j$, similarly for $k_k$ and $k_l$
to avoid double counting. The neutron-proton ($np$) or
the proton-neutron ($pn$) case is more complicated. Here a
free summation over the indices
gives four terms with identical contribution (two for each of
$np$ and $pn$ cases) and a factor of (1/2) would be needed for correct
counting. Thus in our calculation for this case we have confined the
indices $i,~k,~e$ to neutrons and
$j,~l,~f$ to protons with their momentum summations unrestricted.

\section{Results of the s.p. spectrum calculation}
\begin{figure}[h]
\resizebox{2.5in}{3.0in}
{\includegraphics{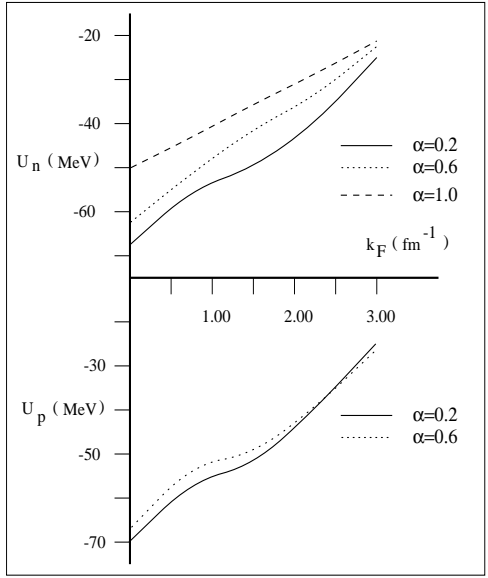}}
\caption{Neutron ($U_n$) and proton ($U_p$) s.p. potentials.}
\label{unup}
\end{figure}
\begin{figure}[h]
\resizebox{!}{2.5in}
{\includegraphics{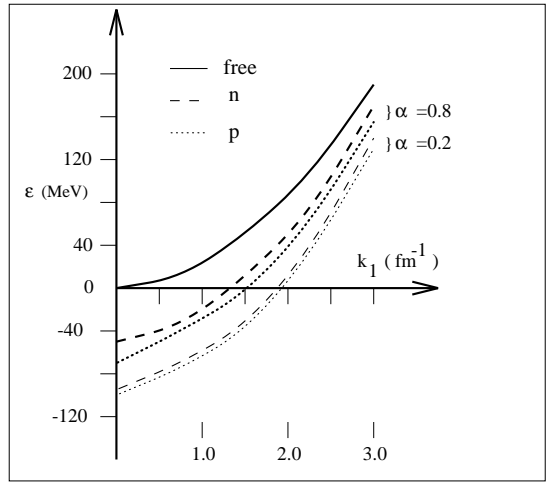}}
\caption{Single particle spectra for neutron and proton calculated 
with the model space approach at $\alp=0.2$ and 0.8 for 
$k_F=1.35$ fm$^{-1}$ with Paris interaction.}
\label{asp2}
\end{figure}
Fig.(\ref{unup}) shows the neutron (upper part) and the 
proton (lower part)
s.p. potentials for different $\alp$. As the proton
fraction decreases the s.p. potentials become less deep and $U_p$
vanishes when the proton fraction becomes zero.

For the asymmetric case as well, the spectrum is continuous
from momentum $0$ to $k_M$ as shown in fig.\ref{asp2},
unlike the usual BHF spectrum which has
a large discontinuity at $k_F$. Beyond $k_M$ we use a free s.p. spectrum
since our method is designed for determination of the s.p. spectrum
within the model space only. Our spectrum has a small discontinuity of
around $4-5$ MeV at $k_M$.

Table 1 lists the $m/m^{*}$ values and zero point energies
at $\alp=0.2$ and $\alp=0.8$ for various average Fermi 
momentum at a model space
boundary of $3.0$ fm$^{-1}$. With increase in $\alp$, $\btu_n$ increases
and $\btu_p$ decreases for small $k_F$.
For the same asymmetry fraction, both $\btu_n$ and $\btu_p$
become deeper with increase in $k_F$.
The change in the effective mass with the asymmetry fraction is very small
though for the same asymmetry fraction it increases with $k_F$.
For $\alp=0.0$, the s.p potential, the effective masses and the
zero point energies match the ones calculated previously for
the symmetric case which also serves as a test for the reliability
of our asymmetric matter calculation.
\begin{table}
\caption{Typical $m^{*}$ and $\btu$  values calculated with the
Paris NN interaction at $\alp=0.2$ and 0.8.}
\begin{center}
\begin{tabular}{l c l c l c}  \hline  \hline
& $k_F$ (fm$^{-1})$ & $m/m^{*}_n$ &  $\btu_n$ (MeV)
& $m/m^{*}_p$ &  $\btu_p$ (MeV)
\\    \hline
$\alp=0.2$ & 1.00 & 1.10 & -29.44 & 1.10 &-31.45 \\
           & 1.20 & 1.15 & -46.61 & 1.16 &-48.89 \\
	   & 1.35 & 1.21 & -61.69 & 1.21 &-63.73 \\
           & 1.50 & 1.26 & -77.38 & 1.26 &-76.83 \\
	   & 1.70 & 1.34 & -96.75 & 1.34 &-97.13 \\  \hline \\

$\alp=0.8$ & 1.00 & 1.08 & -24.66 & 1.10 &-33.05 \\
	   & 1.20 & 1.13 & -40.05 & 1.15 &-49.56 \\
	   & 1.35 & 1.17 & -52.16 & 1.19 &-60.60 \\
	   & 1.50 & 1.23 & -66.47 & 1.24 &-74.05 \\
	   & 1.70 & 1.32 & -86.55 & 1.32 &-91.68 \\  \hline
\end{tabular}
\end{center}
\end{table}
The calculated binding energy for a given combination of the asymmetry
parameter $\alp$ and the Fermi momentum $k_F$ depends on the model-space
size. On minimizing the energy against the model-space momentum, Song, Wang
and Kuo\cite{swk92} found for their MBHF calculations that for each
combination of $\alp$ and $k_F$ there was one value of $k_M$ for which the
energy was a minimum. For the same $\alp$, the energy minimum shifted
towards smaller values of $k_M$. For values of $k_F$ ranging between
$0.50$ to $1.80$ the minimum values of $k_M$ were between $2.80$ to $3.4$.
Based on this we have made our choice of $k_M$ to be $3.0$ fm$^{-1}$.
\section{The RPA equation}
Let us recall our RPA-type secular equation
\beq\label{rpae}
\sum_{e > f} \{ (\eps_i +\eps_j) \del_{ij,ef}+(\bn_i \bn_j -n_in_j) \lam
L_{ijef}(\om) \} Y_{m}(ef, \lam) = \mu_m(\om, \lam) Y_{m}(ij, \lam)
\eeq
with the self consistent condition
\beqar\label{scond}
\mu_m(\om, \lam) &=& \om \equiv \om_{m} ^{-}(\lam) \non \\
L_{ijef}(\om) & \equiv & \bar{G}^{M}_{ijef}(\om)
\eeqar
The above RPA-type equation is in laboratory momentum variables.
As for the symmetric case, we transform the above equation
into its RCM representation.
We also do an angle average for the occupation factor
$(\bn_i \bn_j -n_in_j)=1-(n_i+n_j)$.
We define a function $Q_{R}(k_i,k_j)=1-(n_i+n_j)$.
It is equal to +1 or -1 depending on the values of
$k_i$ and $k_j$.
$Q_{R}$ is $+1$ in regions A and B ($i,j$ both are particles); is
$-1$ in region C ($i,j$ are both holes) and is equal to zero
for all other regions. The value of $Q_{R}$ depends on the angle between
{\bf k} and {\bf K}. Let us denote by $\tau_z$ the $z$-component of the
total isospin $T$ of the two nucleon state i.e., $\tau_z=\tau_i+\tau_j$.
As the two nucleon state can be either $nn$, $pp$ or $np~(pn)$, $\tau_z$
can take the values -1,1 or 0.
For the case when $\tau_z = +1~{\rm or}~ -1$ the two nucleons are identical
and the situation is no different
from the symmetric case ($\bar{Q}_{R}$ is the same as eq. (4.9) in 
Ref. (3).
For $\tau_z=0$ we obtain the angle averaged $\bar{Q}_{R}$ as
\beq\label{rqbar}
\bar{Q}_{R}(k,K) = \left\{ \begin{array}{lr}
 -1 & \mbox{region} \;\; 1 \\
 -|x_{1}| & 2 \\
  1  &  3  \\
 |x_{1}| & 4  \\
 |x_{2}| & 5  \\
 \mbox{min}(|x_{1}|,|x_{2}|) & 6 
 \end{array} \right.
\eeq
\noindent where
\beqas\label{ka}
k_a &=& \frac{k_F^n+k_F^p}{2} \\
x_{1} & =& \frac{k^2 +\frac{1}{4}K^2-k_a^2}{kK}  \\
x_{2} & = & \frac{k_M ^2-k^2 -\frac{1}{4}K^2}{kK}  \\
\eeqas
\noindent The above angle averages are obtained under the
assumption that all values for the angle between {\bf k} and
{\bf K} are equally likely.
\begin{figure}[h]
\resizebox{!}{3in}
{\includegraphics{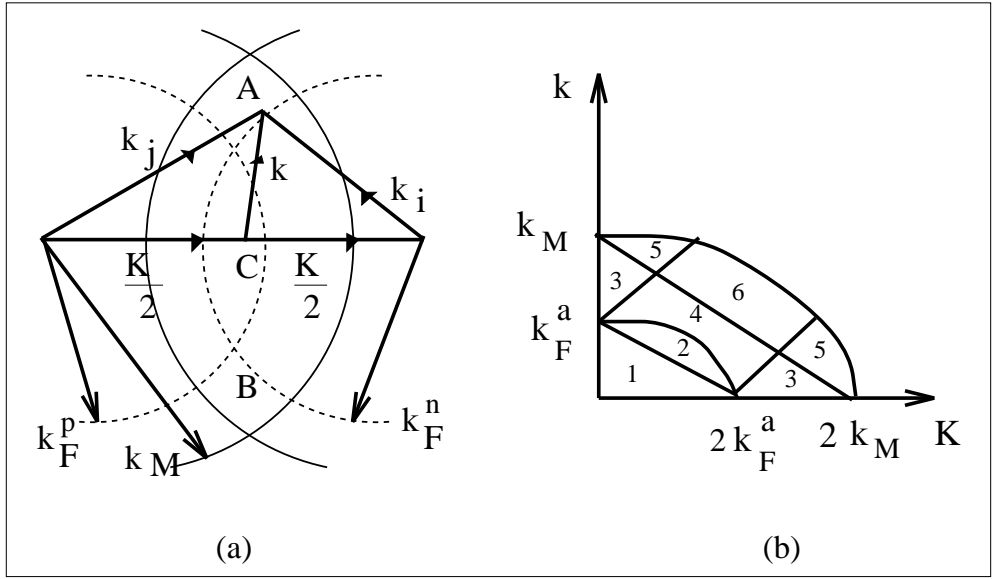}}
\caption{Angle average of occupation factor $Q_{R}(k_{i},k_{j})$}
\label{occupasym.ps}
\end{figure}
The regions 1 to 6 refer to the regions in the $(k,K)$ plane
as shown in the figure.
The replacement of the occupation factor $Q_{R}$ by its angle-averaged
quantity greatly simplifies the RPA-type secular equation we started
with. Now it can be decomposed into separate partial wave channels
\beqar\label{pwrpa}
&& \sum_{\lp} \int d \kp \{ \eps_{kK}~ \del(k-\kp) \del_{l\lp} + \lam
\frac{2 {\kp}^{2}}{\pi} \bar{Q}_{R}(k,K) \lbr kl|L(\om, K)|\kp \lp \rbr \} 
Y_m(\kp \lp k,\lam) \non \\
&& = \mu_m(\om,\lam) Y_m(klK,\lam)
\eeqar
where $\bar{Q}_{R}(k,K)$ is actually
$\bar{Q}_{R}(k,K,k_F^{\tau_i},k_F^{\tau_j},k_M)$. $ \eps_{kK}$ is
the unperturbed energy. There is a subtle point. For the $nn$ or $pp$
cases where the two nucleons are identical
this unperturbed energy may be taken simply as
$\frac{\hb^2}{m}k^2 +\frac{\hb^2}{4m}K^2 +2 U(\bar{k_1},\tau_q) $, where
$\bar{k_1}^2=(k^2 +\frac{1}{4}K^2)$ is the angle averaged value for the momenta
of the two nucleons and $q=n$ or $p$. For the $np~(pn)$ case,
this averaging is more complicated. One way around this is in the
choice of relative $k$ mesh points. For each $K$ mesh point we
choose the $k$ such that the RCM values of $k_n$ and $k_p$ are
either both in the hole region or both in the particle region.
With such a choice of mesh points the sum of the squares of the
individual neutron and proton momenta in the RCM frame, both being either
holes or particles is
$\bar{k}^2_n+\bar{k}^2_p=(k^2 +\frac{1}{4}K^2)$.
Hence the value of the unperturbed
energy is
\bdm\label{e0}
\frac{\hb^2}{m}k^2 +\frac{\hb^2}{4m}K^2+U(\bar{k_1},\tau_n)
+ U(\bar{k_1},\tau_p)
\edm
\noindent with ${\bar{k_1}}^2=(k^2+\frac{1}{4}K^2)/2$.

\noindent
The wave function $(kl)$ denotes the RCM partial wave function
$(klSJ\tau_1 \tau_2,K)$
and similarly for $(\kp \lp)$.
The above equation is to be solved together with the self consistent condition
$\mu_m(\om, \lam)=\om_m^{-}(\lam)$ giving the self-consistent solution
$\om_m^{-}(\lam)$.

The vertex function $L(\om,\tau_1 \tau_2,K)$ in the above 
equation is the irreducible
vertex function which has both one-body and two-body G-matrix 
diagrams\cite{syk87}. 
These contribute significantly to the depletion of s. p. orbits below
$k_F$, especially at high density. 
As we are working in the RCM frame an angle average approximation is 
employed to obtain $L$ in the RCM frame for the $\tau_z=0$ case as
\beqar\label{lrcm}
\lbr kl|L(\om,\tau_n \tau_p,K)| \kp \lp \rbr &=& 
\lbr kl \tau_n \tau_p |\bar{G}^{M}(\om,K w)| 
\kp \lp \tau_n \tau_p \rbr  \non \\
&& +\del_{k \kp}\del_{l\lp} \{\Gam_{\bar{k}_{n}}(\om-\eps_{\bar{k}_{p}},\tau_n)
-U(\bar{k}_{n},\tau_n)  \non \\
&&+\Gam_{\bar{k}_{p}}(\om-\eps_{\bar{k}_{n}},\tau_p)
-U(\bar{k}_{p},\tau_p) \}
\eeqar
\noindent where
\beqar\label{kande}
& &{\bar{k}_n}^2 ={\bar{k}_p}^2 =(k^2+\frac{1}{4}K^2)/2  \non   \\
& & \eps_{\bar{k}_n}=\frac{{\bar{k}_n}^2}{2m} +U(\bar{k}_n,\tau_n)  \non   \\
& & \eps_{\bar{k}_p}=\frac{{\bar{k}_p}^2}{2m} +U(\bar{k}_p,\tau_p)
\eeqar
The other two cases for $\tau_z=-1$ or 1 can be obtained
by suitable substitution of $\tau_p$ and $\tau_n$.
\section{Results and discussions}
\subsection{Binding energy}
The energy shift is given as
\beqar\label{feshift}
\frac{\Del E_0^{pp}}{A} & =& \frac{6}{\pi^2[(k_F^n)^3+(k_F^p)^3]}
\sum_{w} \sum_{\tau_z}
(2J+1) \int_0^1 d \lam~ \int_0^{2k_F^{\tau_z}} dK~ K^2 \non \\
& & \tm \sum_{ml \lp} \int_0^{k_M} dk \; k^2 \int_0^{k_M}
d \kp ~ {\kp}^{2} Y^*_m(klK,\lam)
\tm \lbr kl \tau_1 \tau_2  |G^M(\om, K w) | \kp \lp \tau_1
\tau_2 \rbr Y_m(\kp \lp K,\lam)
\eeqar
\noindent where $\tau_z=-1,0,1$ for $T=1$ and $\tau_z=0$ for $T=0$
and
\beq\label{kf}
k_F^{\tau_z} = \left\{ \begin{array}{ll}
 k_F^n & \mbox{for} \;\;\; \tau_z=-1 \\
 (k_F^n+k_F^p)/2 & \mbox{for} \;\;\; \tau_z=0 \\
 k_F^p & \mbox{for} \;\;\; \tau_z=1
  \end{array} \right.
\eeq
The binding energy per nucleon is given as
\beq\label{be}
\frac{-BE}{A}=\frac{3 {\hb}^2}{20m}[(1+\alp)(k_F^n)^2+(1-\alp)(k_F^p)^2]
+ \frac{\Del E_0^{pp}}{A}
\eeq
\begin{figure}[h]
\resizebox{3in}{3in}
{\includegraphics{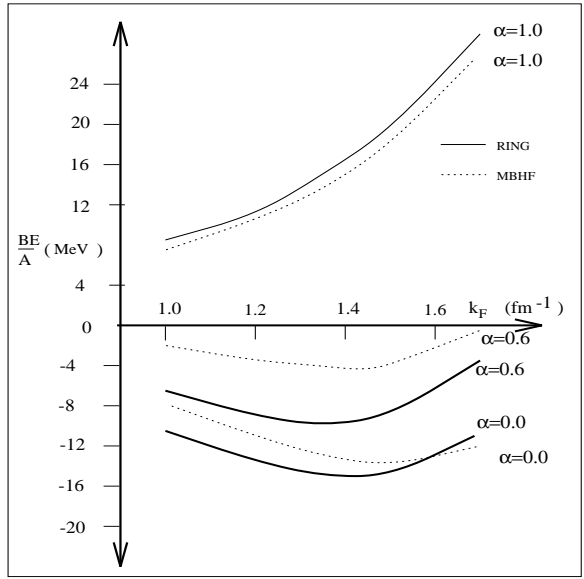}}
\caption{$BE/A$ from Ring diagram summation at three asymmetry 
values : $\alp=0.0,0.6$ and $1.0$}
\label{abe1}
\end{figure}
\begin{figure}[h]
\resizebox{3in}{3in}
{\includegraphics{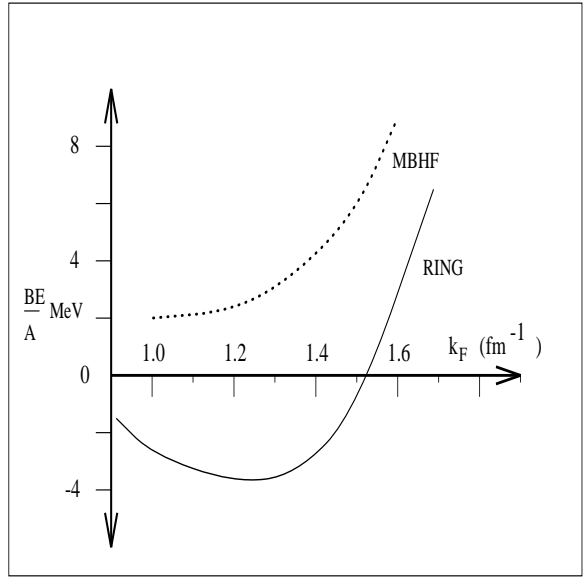}}
\caption{$BE/A$ versus $k_F$ with Ring summation and MBHF 
calculation at $\alp=0.8$. The ring curve shows a saturation 
which is not present in the MBHF curve.}
\label{abe2}
\end{figure}
We present the results of our ring calculation in table 2 and
in figs.\ref{abe1} and \ref{abe2} using Paris NN interactions,
for various values of the Fermi momentum and the asymmetry fraction.
The binding energy at $\alp=0.0$ (symmetric nuclear matter) by our
ring calculation is -15.93 MeV which is good agreement with the
empirical value. The saturation Fermi momentum is 1.43 fm$^{-1}$.
As the asymmetry fraction increases ( the proton fraction decreases),
both the binding energy and the saturation Fermi momentum drop till
at $\alp=1.0$ (zero proton fraction) there is no saturation.
An interesting result of
the ring calculation is the existence of saturation at $\alp=0.8$
which is not present in the MBHF calculation.

\begin{table}
\caption{$BE/A$ in MeV for various combinations of $\alp$ and
$k_F$ (fm$^{-1}$) with $k_M=3.0$ (fm$^{-1}$) and Paris interaction.
Results from (MBHF) calculation are also given for comparison.}
\begin{center}
\begin{tabular}{c c c c c c c}  \hline  \hline   \\
$\alp/k_F$ & 0.65    & 1.00    & 1.20     & 1.35    & 1.50   & 1.70 \\
\\ \hline
0.0        & -6.26   & -10.89  & -14.02   &-15.68   & -15.66 &-12.76 \\
           & (-3.74) & (-8.60) & (-11.40) &(-13.31) &(-14.43)& (-13.47) \\  \\
0.2        & -5.87   & -10.49  & -13.47   &-14.97   & -15.00 &-11.64 \\
           & (-3.42) & (-7.94) & (-10.56) &(-12.38) &(-13.27)& (-12.12) \\  \\
0.4        & -5.78   & -9.60  & -12.32   & -13.30  & -12.76 & -8.58    \\
           &(-2.44)  &(-5.99)  & (-8.00)  & (-9.32) &(-9.60) &(-7.76) \\    \\
0.6        &-5.18    & -7.42   & -9.62    & -10.04  & -8.87  & -2.40 \\
           &(-0.91)  &(-2.68)  &(-3.65)   &(-4.02)  &(-3.30) &(-0.03) \\  \\
0.8        & -1.90   & -3.52   & -3.83    & -3.54   & -1.11  & 6.40 \\
           &(1.29)   &(2.00)   &(2.51)    &(3.52)   &(5.70)  &(11.00) \\  \\
1.0        & 4.03    & 8.11    & 11.03    & 14.22   & 19.02  & 29.65\\
           &(4.08)   &(8.07)   &(10.84)   &(13.74)  &(17.85) &(26.20) \\ \hline
\end{tabular}
\end{center}
\end{table}
\subsection{Symmetry energy}
\begin{figure}[h]
\resizebox{3in}{3in}
{\includegraphics{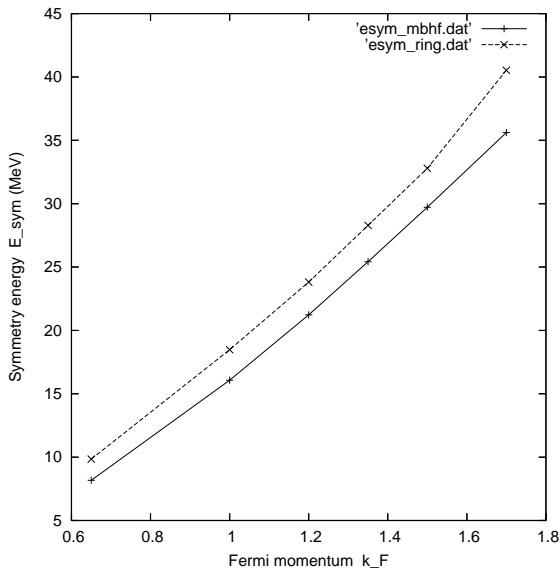}}
\caption{Symmetry energy obtained from Ring (upper curve)
and MBHF (lower curve) calculations.}
\label{esymcomp}
\end{figure}

The symmetry energy is defined as
\beq\label{esym0}
E_{sym}^{Ring}(k_F)=\frac{1}{2}~ \frac{\partial^2 W(\alp,k_F)}{\partial \alp^2}
|_{\alp=0}
\eeq
\noindent where $W(\alp,k_F)$ is the binding energy at a given $\alp$ and
$k_F$.
To ascertain the nature of the dependence of the binding energy 
on the asymmtery parameter we have tried fitting a polynomial
curve of leading order $\alp^2$ and  higher for the residuals 
upto to the order of $\alp ^{6}$ to our data. 
Using eq. (\ref{esym0} and the parameters of the above fit we
obtain the $E_{sym}^{Ring}$ for various values of $k_F$ as shown in table 3.
The corresponding values obtained by the MBHF calculation\cite{swk92}
are also shown for comparison.
The value of the symmetry energy calculated by our method
is consitently higher than the corresponding MBHF calculations as
is shown by fig.\ref{esymcomp}.
Thus the ring diagram summation improves the symmetry energy values as well.
At the saturation density($\rho_0=0.17~ {\rm fm}^{-3}$) $E_{sym}^{Ring}$   
is 28 MeV which is close to the value of 31 Mev reported in an
independent calculation\cite{s74}.
\begin{table}
\caption{$E^{Ring}_{sym}$ calculated from eq. (38) and eq. (39) and
$E_{sym}^{MBHF}$.}
\begin{center}
\begin{tabular}{c l c r}  \hline  \hline  \\
$k_F(fm^{-1})$ & $E^{Ring}_{sym}$ & $E_{sym}^{MBHF}$ & Empirical \\ \\ \hline
0.65           & 9.84             & 8.17             & 10.29      \\
1.00           & 18.48            & 16.07            & 19.00      \\
1.20           & 23.81            & 21.23            & 25.05      \\
1.35           & 28.29            & 25.42            & 29.90    \\
1.50           & 32.79            & 29.72            & 34.68   \\
1.70           & 40.53            & 35.61            & 42.41   \\  \hline
\end{tabular}
\end{center}
\end{table}
Empirically,$E_{sym}^{Ring}$ may be simply evaluated
from the two extreme cases of pure neutron matter and symmetric 
nuclear matter shown in the fourth column of table 3.
\beq\label{esym2}
E_{sym}(k_F)=W(1,k_F)-W(0,k_F)
\eeq
The $E_{sym}$ calculated with eq. (\ref{esym0}) and eq. (\ref{esym2}) differ 
by about .5 MeV to 1.8 MeV .
\subsection{Incompressibility}
In describing the iron-core collapse of a presupernova using hydrodynamical
models, the basic physical inputs are the initial mass of the core and the
nuclear EOS. The EOS of asymmetric nuclear matter exhibits a minimum which
disappears before pure neutron matter is reached. Therefore we expect that
the incompressibility modulus decreases with respect to the symmetric case
and vanishes before the proton fraction vanishes.
The incompressibility modulus is defined as
\beq\label{kap}
\kap_0(\alp)=k_{F_{0}}^2(\alp)~\frac{d^2 W(\alp,k_F)}{dk_F^2}~
|_{k_F=k_{F_{0}}(\alp)}
\eeq
\noindent where $k_{F_{0}}(\alp)$ is the saturation Fermi momentum at the
given $\alp$.
One of the most sophisticated investigation of the $\alp$ dependence of 
$\kap_0$ and
$\rho_0$ so far has been done in Ref. (14) in the framework of 
BBG (Brueckner-Bethe-
Goldstone) theory. They found that
for low $\alp$ values ($\alp \leq 0.4$) $\kap_0$ and $\rho_0$
show a parabolic dependence on $\alp$
\beqar\label{kappa}
\kap_0(\alp) &=& \kap_0(0)(1-a~\alp^2)  \\
\rho_0(\alp) &=& \rho_0(0)(1-b~ \alp^2)
\eeqar
\noindent with $\kap_0(0)=185$ MeV, $a=2.027$ and $\rho_0(0)=0.289$ fm$^{-3}$,
$b=1.115$.
In table 4 we give the $\kap_0$ values obtained from the ring calculation.
Our $\kap_0$ values also seem to obey a parabolic dependence on $\alp$ but with
$a=1.21$ and $\kap_0(0)=112$ MeV.
We may note here that the values of $\kap_0(0)$ and $a$ in Ref. (14)
were obtained by means of a least square polynomial fit to the BHF values
of the binding energy and the values are quite sensitive to the degree
of the polynomial used for the fit.
\begin{table}
\caption{$\kap_{0}(\alp)$ calculated from eq. (40).}
\begin{center}
\begin{tabular}{l c r r}  \hline \hline
$\alp$ & $k_F^{sat}(fm^{-1})$ & $\kap_0(\alp)$ &  $W_0(\alp)$
\\ \hline
0.0    & 1.43                 & 114.51           & -15.93  \\
0.2    & 1.43                 & 100.20           & -15.25  \\
0.4    & 1.39                 & 92.74            & -13.33  \\
0.6    & 1.33                 & 67.35            & -10.05  \\
0.8    & 1.24                 & 22.95            & -3.85   \\  \hline
\end{tabular}
\end{center}
\end{table}

Our value of $\kap_0(0)=112$ MeV is in good agreement with the one
suggested by
Brown and Osnes\cite{bo85} for symmetric nuclear matter.
At $\alp=0.33$ the value of $\kap_0$
from our fit is 97 MeV which is close to the empirical value of 90 MeV
used by Baron, Cooperstein and Kahana\cite{bck85}
to get the maximum explosion energy in their hydrodynamical calculations.
From their numerical investigation the authors of Ref. (8) concluded that
the softening of the EOS plays a crucial role in generating 
prompt explosion for
stars in the mass range of $12-15~M_{\odot}$ (where $M_{\odot}\sim 2 
\tm 10^{23}~ g$
is the sun's  mass). This conclusion has been questioned
by more refined calculations\cite{b85,mb89,bc90}. However, even in models
where the direct explosion mechanism fails, a softer EOS is helpful to the
shock. The decrease in incompressibility with increase in $\alp$ is quite
intuitive when we consider that going from bound symmetric matter to unbound
neutron matter, the minimum of the binding energy gradually disappears. 
Therefore
from eq. (40) the nuclear incompressibility decreases and vanishes for a
certain value of $\alp$.

\section*{Conclusion}
In conclusion, a fully microscopic calculation has been done using 
the ring-diagram 
summation for the EOS of asymmetric nuclear matter. The numerical computation
has been done with both the Paris\cite{parispot} and the Bonn\cite{bonnpot} 
potentials and the results
are in satisfactory agreement with the empirical values. The model-space 
size is treated as a parameter. The inclusion of ring-diagrams 
increases the role of tensor forces and both the binding energy and the 
saturation density values
are lower and closer to the empirical values than those obtained with
previous calculations. The symmetry energy values also show an improvement.
and is in better agreement with other independent calculations.
 The derived incompressibility exhibits a parabolic
dependence on the asymmetry fraction, in qualitative agreement with the
empirical asymmetry dependence used in the literature. 

The ring diagram approach employs an infinite order summation of
particle-particle hole-hole ring diagrams. The infinite oder summation 
technique of Yang,Heyer and Kuo\cite{yhk86} is applicable to particle-hole
ring diagrams as well. But these diagrams have not been included in nuclear
matter calculations. There is reason to believe that the effect of the
particle-hole ring diagrams on binding energy calculatins is not 
very appreciable and they are less
important than the particle-particle ring diagrams. The lowest-order ring
diagram is the pphh diagram of first order in $G^M$. The second-order
diagram may be taken as either pphh or ph and is second order in $G^M$.
Thus the contribution to the ground state energy shift from the
ph diagrams comes from the three vertex diagram which is third order in
$G^M$ (fig. 1(d)). Studies\cite{dfm82,mbook,day81} have indicated that 
the particle-hole diagrams
converge rapidly and they may not be impotant in nuclear matterbinding 
energy calculations. This view is also supported by some Lipkin model 
calculations\cite{yhk86}.

As noted earlier, an interesting result of our
ring calculation is the existence of saturation at $\alp=0.8$
which is not present in the MBHF calculation. This behaviour would be
of relevance in astrophysical systems which are essentially neutron rich.
It would interesting to study the neutron star properties with this method.
To do that one needs to extend the present calculation to higher densities in 
a relativistic frame work.

The authors would like to thank Prof. G. E. Brown for his support
and encouragement throughout the course of this work.


\end{document}